\documentclass[12pt,a4paper,final]{iopart}
\newcommand{\be}{\begin{equation}}
\newcommand{\ee}{\end{equation}}
\newcommand{\bea}{\begin{eqnarray}}
\newcommand{\eea}{\end{eqnarray}}
\newcommand{\ba}[1]{\begin{array}{#1}}
\newcommand{\ea}{\end{array}}
\usepackage{amssymb}
\usepackage{mathrsfs}
\usepackage{color}
\usepackage{iopams} 
\usepackage{graphicx}
\usepackage[breaklinks=true,colorlinks=true,linkcolor=blue,urlcolor=blue,citecolor=blue]{hyperref}

\begin{document}
\setlength{\topmargin}{0.2in}

\title{Formation of a molecular ion by photoassociative Raman processes }
\author{Dibyendu Sardar$^1$, Somnath Naskar$^{1,3}$, Arpita Pal$^1$, 
 Hamid Berriche$^{4,5}$ and Bimalendu Deb$^{1,2}$}
\address{$^1$Department of Materials Science, Indian Association for the Cultivation of Science (IACS), Jadavpur, Kolkata 700032, INDIA.
\\ $^2$Raman Center for
Atomic, Molecular and Optical Sciences, IACS.  \\ $^3$Department of Physics, Jogesh Chandra Chaudhuri College, Kolkata-700033, India, \\
$^4$Laboratory of Interfaces and Advanced Materials, Physics Department, Faculty of Science, 
 University of Monastir, 5019, Tunisia. \\
 $^5$Department of Mathematics and Natural Sciences, School of Arts and Sciences, American University of RAK, P. O . Box 10021, RAK, UAE.}

\begin{abstract}
We show theoretically that it is possible to form  a cold molecular ion from a pair of colliding 
atom and ion at low energy by photoassociative two-photon Raman
processes. We explore the possibility of stimulated Raman adiabatic passage (STIRAP) 
from the continuum of ion-atom scattering states to an ionic molecular state.  We provide physical conditions under which coherent population transfer is possible in stimulated Raman photoassociation. Our
results are important for experimental realization of PA in ion-atom cold collisions.
\end{abstract}


\maketitle

\section{Introduction}

In recent years, a number of schemes has been used to synthesize cold molecules from cold atoms in the presence
of external fields. One of the most important methods in this regard is the photoassociation (PA)
of cold atoms \cite{weiner1999,jones2006,band1995,fioretti1998}. One- or two-color PA has been successfully employed  to produce neutral cold molecules in excited or ground-state electronic potentials, respectively. Another important method is magnetoassociation \cite{ni2008,julienne} by which exotic  Feshbach molecules \cite{fes_molecule,fes_molecule1,fes_molecule2} are produced in the presence of an external magnetic field. Although the formation of cold molecular ion by PA between a cold atom and a cold ion had been theoretically predicted \cite{rakshit2011} and analysed \cite{dulieu2011} over the last few years, the experimental realization of ion-atom PA is yet to be achieved. The difficulty in experimental ion-atom PA mainly stems from the fact that it is difficult to obtain sufficiently low temperature (down to sub-milliKelvin) in a hybrid ion-atom system \cite{markov_hbd,goodman,zipkes,zipkes1,grier}
 as 
required for ion-atom PA process. As ions are usually trapped and cooled by radio-frequency 
fields, the major hindrance in cooling
trapped ions arises from the trap-induced micro-motion of the ions. But this difficulty can be overcome by using
optical traps \cite{huber2014} where both ions and atoms can be trapped and cooled. Experimental progress towards optical trapping of ions
is underway \cite{nature-photon,schneider2012,oplatice,oplatice1,oplatice2} and so one can expect that ion-atom PA will soon be realized in an optical trap. 

In this paper, we present and analyse a theoretical method  for creating cold molecular ion from a colliding pair of alkali ion and alkali atom by photoassociative Raman processes.
Our model is schematically shown in FIG.\ref{fig1}. For illustration, we consider (LiCs)$^+$ system, but our theoretical formulation is most general and can be applied to any other ion-atom system. 
We consider molecular level structure in $\Lambda$-type configuration involving one continuum of ion-atom scattering states and two molecular bound levels. As depicted in the inset of FIG. \ref{fig1}, the continuum of states $\mid E \rangle$ 
in the electronic ground-state manifold of the ion-atom system is coupled to the excited ro-vibrational bound state $\mid 2 \rangle$ in the electronic potential $V_2(r)$ by
PA laser $L_2$ and $\mid 2 \rangle$ is coupled to the ro-vibrational bound state $\mid 1 \rangle$ in a lower potential $V_1(r)$ by 
laser $L_1$. The resulting configuration looks like a three-level $\Lambda$ system with one of the lower ground-state sub-levels being replaced by the continuum. Here, we are interested in transferring population from $\mid E \rangle $ to $\mid 1 \rangle $ via the intermediate state $\mid 2 \rangle$ by a two-photon Raman process. One can consider two possible processes - incoherent and coherent Raman processes to accomplish the state transfer. In the incoherent process, one laser photon from $L_2$ excites the system to make an upward transition $\mid E \rangle \rightarrow \mid 2 \rangle$ followed by bound-bound downward transition $\mid 2 \rangle \rightarrow \mid 1 \rangle$ by spontaneous emission of a photon. The net result is the formation of the bound state $\mid 1 \rangle$. The success of such incoherent Raman process in molecular systems relies on the availability of favorable Franck-Condon (FC) overlap between the continuum and the bound state and also between the two bound states. Nevertheless, the 
efficiency of transfer is in general limited in the case of incoherent Raman processes. In a recent paper C\^{o}t\'{e} and co-workers \cite{kote2016} have theoretically discussed the possibility of forming molecular ion by incoherent Raman photoassociation. 

 An efficient way for state transfer is the coherent method of stimulated  Raman process using two laser pulses. When the two slowly varying time-dependent laser pulses are applied  in a counter-intuitive way, that is, if laser $L_1$ 
is applied first between the target ($\mid 1 \rangle$) and the intermediate state ($\mid 2 \rangle$) which are initially empty, and then after a time delay the laser $L_2$ is applied between the initial and the intermediate state,  the resultant process of state transfer is popularly known as stimulated Raman adiabatic process (STIRAP) \cite{gaubatz1990,kuhn1992,halfmann1996,martin,kral2007,cbb_bbc_bergmann} subject to the fulfillment of an adiabatic condition. In the case of discrete three-level $\Lambda$ systems driven  by two Raman lasers, associated with a host of coherent processes including STIRAP is the existence of a dark state (DS) which is a special eigenstate of the system. The 
coupling  of the excited intermediate state with the DS is zero. An adiabatic evolution of the DS can be executed by slowly varying the intensity of the applied lasers in
 a proper sequence, allowing nearly loss-less population transfer between  two ground-state sub-levels.

 The possibility of STIRAP via an intermediate continuum of states is discussed by different groups $\cite{car1992,car1993,nakajima1994,unanyan1998}$. Besides, taking continuum as an initial or final target state, state transfer is theoretically studied $\cite{opticexprs1999,cbb_bbc_mshapiro}$. STIRAP in a PA system is a subject of such category of state transfer. Indeed, the possibility of STIRAP in such systems is a matter of intense theoretical debate $\cite{java1998,mackie1999,vardi2002,java2002}$. Photoassociative adiabatic passage (PAP) in ultra-cold rubidium  atomic gas has been studied $\cite{cbb_bbc_eashapiro,err_eashapiro}$. In 2005, P. D.  Lett and coworkers 
experimentally observed EIT-like spectral features in two-color PA of ultra-cold sodium $\cite{eit_pa_expt_lett}$. 
Experimental signatures of a dark resonance has also been found in ultra-cold meta-stable helium 
by Cohen-Tannoudji's group $\cite{ds_pa_expt_cohen1}$ and in ultra-cold strontium atoms by Killian and co-workers $\cite{ds_pa_expt_killian}$. Although, these findings indicate to the 
possibility of STIRAP from cold atomic gas to cold molecules, such process is not yet clearly established theoretically or experimentally. 
 
In this paper, we adopt the Fano diagonalisation method \cite{fano1961} to obtain a dressed continuum state and find the physical conditions under which the dressed state behaves approximately as a DS. We then 
     explore the regime where the DS can be adiabatically evolved to perform a STIRAP-like process to transfer the atomic population from scattering continuum to molecular population in electronic ground-state. We use a 
     strongly dipole allowed PA transition to an excited molecular
ionic state which, at long separation, corresponds to an ion in $s$-state and a neutral atom in $p$-state. This excited state is 
short-lived. Here we use this molecular state as an intermediate state in performing STIRAP.
We use two laser pulses in a counter-intuitive way. We have shown that 
effective coherent population transfer is possible.

        This paper is organized in the following way. In Sec.II, we describe our theoretical model and analyse the condition for formation of DS. Numerical results are presented and discussed in Sec.III. In last section we make some concluding remarks.        

\section{The theoretical formulation}
\subsection{The model}

\begin{figure}
 \includegraphics[width=0.9\linewidth]{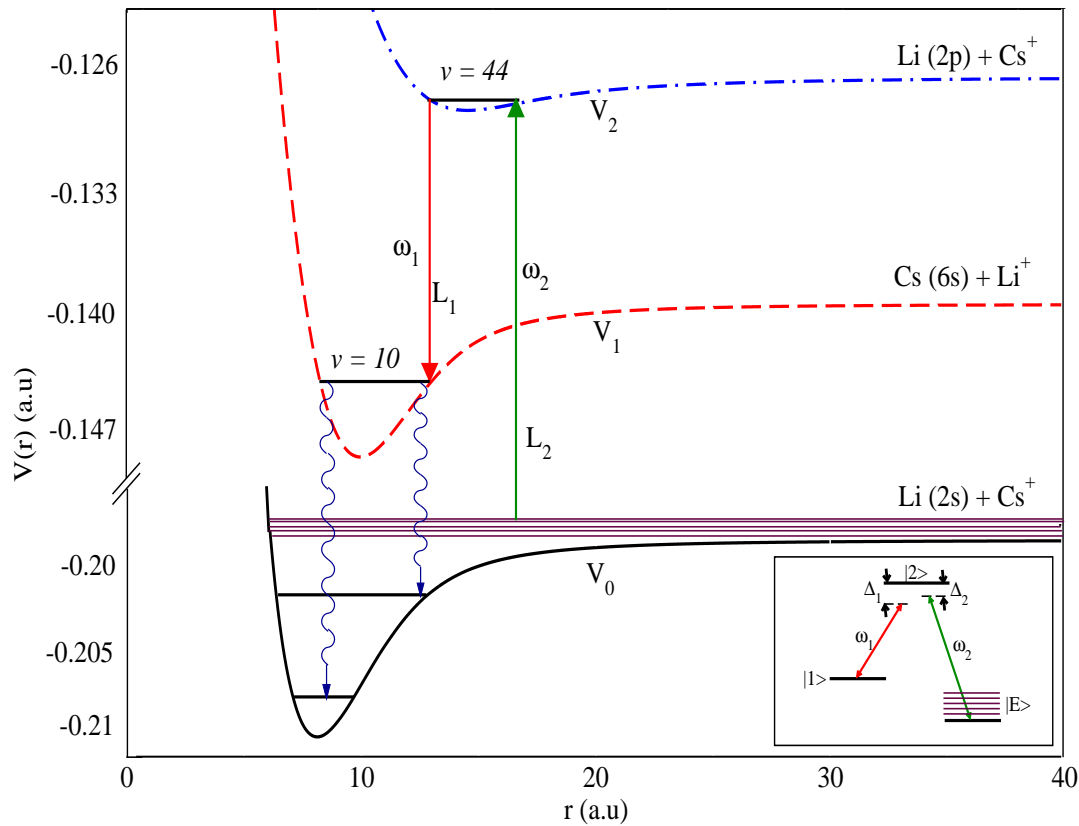}
 \caption{Adiabatic potential energy curves of the three lowest electronic states of (LiCs)$^+$ ion. The
 asymptotic states of separated ions and atoms are indicated. Three states can be considered like a $\Lambda$ type 
 system as indicated above. The band of horizontal lines above the threshold of the lowest potential
 indicate the continuum of states of low energy collision between the closed-shell Cs$^+$ ion and
 ground-state Li atom. The inset shows a three-level $\Lambda$-like structure with one lower ground-state sub-level in discrete $\Lambda$-system being replaced by the continuum. Three potentials are in $ ^2\Sigma^+$ electronic state.}
 \label{fig1}
\end{figure}

Here we present our theoretical formulation using a model system of (LICs)$^+$ as depicted in FIG.\ref{fig1}. The adiabatic potential data of (LICs)$^+$ system are calculated using pseudopotential method and described elsewhere \cite{arpita2016}.
Initially the system is in the continuum states $\mid E \rangle $  of the ground-state adiabatic potential 
$V_0(r)$. The potential $V_0(r)$ asymptotically corresponds to the heavier alkali element in the ionic 
state (Cs$^+$) and the lighter one in the charge-neutral state (Li). The potential $V_1(r)$ asymptotically corresponds to the charge-exchanged state of that of $V_0(r)$. For both $V_0(r)$ and $V_1(r)$ potentials,
the ion and the neutral atom are in electronic ground-state ($S$) at long separations. Therefore, 
 the molecular electric  dipole coupling between the electronic states of these 
two potentials vanishes at large separations while it is finite but small at short separations only. As a result, the ro-vibrational states 
supported by $V_1(r)$ are expected to be long-lived or meta-stable. However, accessing these states via free-bound transitions from the continuum of $V_0(r)$   
is extremely difficult due to the poor Franck-Condon (FC) overlap integral at short separations. In contrast, the transition dipole moment between the electronic states of $V_0(r)$ and $V_2(r)$ 
is large and constant at large separations where the potential $V_2(r)$ correspond to the lighter element in the excited $P$ state while the heavier one remains 
in the ground ionic state as in the case of $V_0(r)$. Therefore, the electric dipole transition between the electronic states of $V_0(r)$ and $V_2(r)$ 
is similar to the strongly allowed electric dipole $E1$ transition. On the other hand, the ro-vibrational states in $V_2(r)$ are short-lived. So, from the practical point of view, it will be difficult to produce or probe the molecular states in $V_2(r)$ by photoassociation. Our primary purpose here is to form a meta-stable bound state in $\mid1\rangle$ in the potential $V_1(r)$ via two-photon coherent Raman process using the lossy state $\mid2\rangle$ as an intermediate state.
 Once a molecular ion is formed in the potential $V_1(r)$, the molecular ion in the ground-state potential $V_0(r)$ can be created either by spontaneous or stimulated bound-bound emission. Interestingly, 
 since the ro-vibrational state in $V_1(r)$ is meta-stable, population inversion between two molecular levels in $V_0(r)$ and $V_1(r)$ may be achieved.  At low temperatures (milli or sub-milliKelvin regimes), depending on the rotational quantum number of the excited molecular state that is coupled by PA laser, only one or two or a few partial waves of the ion-atom scattering state can contribute to the free-bound FC factor. This allows us to select a narrow range of energy for continuum-bound coupling as elaborated in the next section.

 For the particular system (LiCs)$^+$ under consideration, we find that vibrational state $v_2 = 44$ in $V_2(r)$ has significant FC 
factor with the continuum states of $V_0 (r)$ at energies ranging from microKelvin to milliKelvin regime.
We notice
that the inner turning point of $v_2=44$ lies at almost same separation of the outer turning point of vibrational state $v_1 = 10$ in $V_1 (r)$ potential. This means that, according to FC principle, these two vibrational states are the most useful or favorable for our purpose.
The state $\mid1\rangle$ is coupled to $\mid2\rangle$
 via laser pulse $L_1$ with frequency $\omega_1$ and electric field ${\mathbf E}_{L1} $.
$\mid2\rangle$ is coupled to the continuum of states $\mid E\rangle$ via laser pulse $L_2$ having frequency $\omega_2$ 
and field ${\mathbf E}_{L2} $. We denote $E_{v_1}$ ($E_{v_2}$) as the energy of the bound state $\mid1\rangle$ ($\mid2\rangle$).
The Hamiltonian of the system under rotating wave approximation (RWA) can be expressed as $H = H_0 + H_I$, where 
\begin{eqnarray}
\label{h0}
 H_0 &= &\hbar(\Delta_1 - \Delta_2)\mid 1\rangle\langle 1\mid + (-\hbar\Delta_2\mid 2\rangle\langle2\mid ) + \int E'\mid E'\rangle\langle E'\mid dE' 
\end{eqnarray}
is the free part and 
\begin{eqnarray} 
\label{hi}
 H_I = \int\Lambda_{E'2}\mid E'\rangle\langle2\mid dE' +      
        \hbar G_1 \mid 1\rangle\langle2\mid  + {\rm H.c.} 
\end{eqnarray} 
is the interaction part of the Hamiltonian. Here 
 $G_1=\frac{1}{\hbar}\langle1\mid {\bf D.{ E}}_{L1}\mid 2\rangle$ is the bound-bound Rabi frequency and
$\Lambda_{2E}=\langle2\mid {\bf D.{ E}}_{L2}\mid  E\rangle$ is the free-bound coupling parameter, $\textbf{D}$ being the molecular
electric dipole moment. The detuning parameters introduced above are defined as $\Delta_1 = \omega_1 - (E_{v_2} - E_{v_1})/\hbar  $ 
and $\Delta_2 = \omega_2 - E_{v_2}/\hbar $. The limits of the energy integrals are [$0,\infty$] and remain the same throughout the paper.
Following the Fano diagonalisation method the dressed eigen state of $H$ can be written as
 \begin{eqnarray}
\mid E\rangle_{dr} &=& B_{1E} \mid 1\rangle + B_{2E} \mid 2\rangle + \int F_E(E') \mid E'\rangle
\label{drs_state}                                  dE'
\end{eqnarray}
where $B_{1E}$, $B_{2E}$, $F_E(E')$ are expansion coefficients as defined in the Appendix.
Since $\mid E\rangle_{dr}$ is energy normalized we must have $\int_{dr} \langle E'\mid E\rangle_{dr} dE=1$ which gives 
\begin{eqnarray}
\hspace{-2cm}\int \mid B_{1E}\mid ^2 dE+\int \mid B_{2E}\mid ^2 dE+\int \int \mid F_E(E'')\mid ^2dE dE''=P_1+P_2+P_C=1
\end{eqnarray}
where $P_1$, $P_2$ and $P_C$ denote the probability that an atom-pair is in the state $\mid 1\rangle$,  $\mid 2\rangle$ and the continuum, respectively.

\subsection{Dark state condition}

As stated earlier, for the systems involving a continuum, the existence of a DS is not obvious, and perhaps an exact DS is not possible. 
We search both analytically and numerically for a condition for which the contribution of the excited intermediate state to the dressed state Eq.(\ref{drs_state}) is small
for a wide range of collisional energy. The purpose is to use the dressed state continuum as an approximate dark state that can be used for reasonably efficient STIRAP. 
The expression for $B_{2E}$ (for derivation, see Appendix) is given by 
\begin{eqnarray}
\hspace{-2cm} B_{2E} =\frac{\left[E+\hbar(\Delta_2-\Delta_1)\right]\left[E+\hbar \Delta_2 -E^{sh}_1 -i\hbar\Gamma_{2E}/2\right] \Lambda_{2E}}{\left[E+\hbar(\Delta_2-\Delta_1)-E^{sh}_2\right]\left[(E+\hbar\Delta_2-E^{sh}_1)^2+(\hbar\Gamma_{2E}/2)^2\right] +i\hbar^3\Gamma_{2E} G^2_1/2 }.                 
\end{eqnarray}
Clearly, this cannot be zero for all energies. For a particular set of $\Delta_1$ and $\Delta_2$ (such that 
$\Delta_1> \Delta_2$), the exact two-photon resonance condition is satisfied for a particular energy $E=\hbar\left(\Delta_1- \Delta_2\right)$.
Let us denote this energy as $\bar {E} = \hbar (\Delta_1 - \Delta_2)$.
In case of a discrete three-level $\Lambda$ system, DS condition is exactly fulfilled only at two-photon resonance.
For our model with one continuum, the two-photon resonance condition has changed due to continuously
varying energy of one of the lower state. Since
$E \ge 0$, the two-photon resonance condition for $\bar{E}$ can only be fulfilled if $\Delta_1 > \Delta_2$, implying
that the bound-bound one-photon detuning $\Delta_1$ must be greater than the one-photon continuum-bound detuning $\Delta_2$
which is defined as $\Delta_2 = \omega_2 - (E_{v_2} - E_{th})/\hbar$, where $E_{th}$ is the threshold value of the 
open (continuum) channel. 

Thus for $E = \bar{E} = \hbar\left(\Delta_1- \Delta_2\right)$ the coefficient $B_{2E}$ vanishes. As a result, the  dressed continuum of Eq.(\ref{drs_state}) reduces to the form 
\bea 
\mid \bar{E} \rangle_{{\rm dr}} = \bar{B}_{1}\mid 1\rangle 
 + \int \bar{F}_E(E')\mid E'\rangle dE'
\eea 
If the coupling of this state with the excited state $\mid 2 \rangle$ vanishes then we can regard this as a DS. This means that 
\bea 
\langle 2 \mid H_I \mid \bar{E} \rangle = 0 
\eea 
This condition may be fulfilled by adjusting the phase of the two lasers. 
So, a DS for the entire range of continuum energy cannot be achievable. However, if we consider that the system is
kept at very low temperature so that the kinetic energy distribution of the free atoms are very narrow about
$\bar E$ then most of the atoms lie very near to $\bar{E}$. 

Assuming the condition as mentioned above is prevailed, we define the DS condition as $P_2$ = $\int \mid B_{2E}\mid ^2 dE << 1$  as this would
lead to small contribution of $\mid 2\rangle$ to the dressed state. As shown in the Appendix, the expressions for $P_1$ and $P_2$ are 
\begin{eqnarray}
\label{p1}
 P_1&=& \frac{\hbar^3}{2} |G_1|^2\int Z^{-1}\Gamma_{2E}(E+\hbar \Delta_2-E^{sh}_1)^2 dE\nonumber \\
  &+&\frac{\hbar^5}{8} |G_1|^2\int Z^{-1} \Gamma_{2E}^3 \hspace{0.1 cm}  dE 
\end{eqnarray}
and
\begin{eqnarray}
\label{p2}
P_2&=&\frac{\hbar}{2}\int  Z^{-1}  [E+\hbar(\Delta_2-\Delta_1)]^2 (E+\hbar \Delta_2 -E^{sh}_1)^2\Gamma_{2E}dE \nonumber \\ 
                                       &+& \frac{\hbar^3}{8} \int Z^{-1}[E+\hbar(\Delta_2-\Delta_1)]^2 \Gamma_{2E}^3 dE  
\end{eqnarray}
where
\begin{eqnarray}
 Z &=& [E+\hbar(\Delta_2-\Delta_1)-E^{sh}_2]^2[(E+\hbar\Delta_2-E^{sh}_1)^2+(\hbar\Gamma_{2E}/2)^2]^2 \nonumber \\
        &+&\hbar^6\Gamma^2_{2E} G^4_1/4] 
\end{eqnarray}
$E^{sh}_1$ and $E^{sh}_2$ are shift quantities whose expressions are given in the Appendix. These quantities are usually small and can be neglected. We consider that the energy-dependence
of the free-bound coupling is very weak and can be assumed to be constant which is calculated at some energy in the energy regime of interest near $\bar E$. This approximation is commonly
known as slowly varying continuum approximation (SVCA) $\cite{svca}$. Now, for computational convenience we scale all the energy terms of Eq.(\ref{p1}) and Eq.(\ref{p2}) by $G_1$ and rewrite them as  
\begin{eqnarray}
P_1 =\int  \frac{[(\epsilon +\delta_2)^2 +\beta^2]\beta }{(\epsilon +\delta_{12})^2+[(\epsilon +\delta_2)^2+\beta^2]^2 +\beta^2}d\epsilon  
\end{eqnarray}
and
\begin{eqnarray}
P_2=\int  \frac{(\epsilon + \delta_{12})^2[(\epsilon + \delta_2)^2 +\beta^2]\beta }{(\epsilon + \delta_{12})^2[(\epsilon + \delta_2)^2 +\beta^2]^2 + \beta^2}d\epsilon
\end{eqnarray}
where  $\delta_{12}=\hbar(\Delta_2-\Delta_1)/G_1$, $\delta_2=\hbar\Delta_2/G_1$, $\beta=|\Lambda_{2\bar E}| ^2/\hbar G_1$. So, in terms of scaled parameters, the two-photon
resonance condition is $\bar \epsilon + \delta_2 = 0$ where $\epsilon=E/G_1$. Now, for the $\epsilon \ne \bar \epsilon$, the system will be off-resonant and as a consequence there is no DS.  

\subsection{Adiabatic regime}

In order to develope an intuitive understanding how the well-known discrete $\Lambda$ configuration can be compared with $\Lambda$-like configuration with a continuum as far as DS and its application to STIRAP are concerned, we first recall the DS condition in the discrete $\Lambda$ system.
If the continuum state $\mid E\rangle$ in the inset of FIG.\ref{fig1} is replaced by a bound state $ \mid 3\rangle$, it will be a three-level discrete $\Lambda$ configuration for which the total Hamiltonian 
under RWA can be written as 3$\times$3 matrix. On diagonalising the matrix, one obtains three eigen states $\cite{gaubatz1990,kuklinski1989}$ 
\begin{eqnarray}
 \mid a^0\rangle = \cos \theta \mid 3\rangle - \sin \theta \mid 1\rangle  
\end{eqnarray}
\begin{eqnarray}
 \mid a^+\rangle = \sin\phi \sin\theta \mid 3\rangle + \cos \phi\mid 2\rangle + \sin \phi\cos\theta \mid 1\rangle  
\end{eqnarray}
\begin{eqnarray}
 \mid a^-\rangle = \cos\phi \sin\theta \mid 3\rangle - \sin \phi\mid 2\rangle + \cos \phi\cos\theta \mid 1\rangle  
\end{eqnarray}
with eigen value $\alpha_0 = \Delta_2$, $\alpha_\pm =\frac{1}{2} [\Delta_2  \pm (\Delta^2_2 + G^2_1 + G^2_2)^{1/2}]$ respectively. The continuum-bound coupling is replaced by another bound-bound coupling parameter $G_2$.
The angle $\theta$ and $\phi$ are defined as 
\begin{eqnarray}
 \tan \theta = \frac{G_2}{G_1} 
\end{eqnarray}
\begin{eqnarray}
 \tan \phi = \frac{(G_1^2 + G_2^2)^{1/2}}{(\Delta^2_2 +G_1^2 + G_2^2)^{1/2} - \Delta_2} 
\end{eqnarray}

The degree of slowness or adiabaticity is assured when the rate of nonadiabatic coupling is small compared
to the separation of the corresponding eigenvalues. i.e
\begin{eqnarray}
   \Bigg|\Bigg\langle a^\pm\Bigg|\frac{d}{dt}\Bigg|a^0\Bigg\rangle\Bigg|^2 \ll |\alpha_0 - \alpha_\pm|^2
\end{eqnarray}
which implies $\cite{gaubatz1990,kuklinski1989}$
\begin{eqnarray}
 \Bigg|\Bigg\langle a^+\Bigg|\frac{d}{dt}\Bigg|a^0\Bigg\rangle\Bigg|^2 + \Bigg|\Bigg\langle a^-\Bigg|\frac{d}{dt}\Bigg|a^0\Bigg\rangle\Bigg|^2 = \Bigg|\frac{d\theta}{dt}\Bigg|^2
\end{eqnarray}
Thus, the adiabaticity is maintained when the rate of change of the mixing angle ($\theta$) is sufficiently small, i.e when
\begin{eqnarray}
 |\dot \theta| \ll |\alpha_0 - \alpha_\pm|
\end{eqnarray}

In the case of $\Lambda$-like configuration with a continuum, as mentioned earlier, an approximate DS like situation can be achieved at ultra-cold temperature. We can set $\bar E$ to the most probable kinetic energy  of a thermal distribution by adjusting $\Delta_1-\Delta_2$. Since $T$ is a measure of spread of the thermal energy distribution, all atoms having energy within $\bar E-k_{b}T/2$ to $\bar E+k_{b}T/2$ can be approximately said to be in a DS, $k_{b}$ being the Boltzmann constant.
Eventually, for this case, the energy difference between a DS and a bright state is $\sim k_{b}T$. Thus, the adiabaticity condition can be written as 
\begin{eqnarray}
\label{adia}
 \left|\frac{d\beta}{dt}\right|\ll k_{b}T/\hbar 
\end{eqnarray}
 The condition Eq.(\ref{adia}) should be maintained in conformity with the energy- or temperature-dependence of the FC factor and $\bar E = 2k_{b}T = \hbar(\Delta_1 - \Delta_2)$. The free-bound coupling or free-bound FC factor is only significant at low energy. This means $T=\frac{\hbar}{2k_{b}}\left(\Delta_1-\Delta_2\right)$ should be small enough so that FC factor is large. However, as we discuss in the next section, the rotational selection rule allows us to restrict the energy range for effective free-bound coupling.

\section{Results and discussion}

To select the two most suitable bound states of (LiCs)$^+$ system for our purpose, 
we first calculate the scattering-state and  a large number of bound state wave functions by a standard 
re-normalized Numerov-Cooley method $\cite{jonson}$. We have calculated the deeply bound states as well as 
bound states close to continuum of both the $V_1(r)$ and $V_2(r)$ potentials. To search for appropriate bound states for building our model, 
\begin{table}[ht]
 \caption{Rabi-frequency calculated between vibrational level $v_2 = 44$ and different  vibrational levels $v_1$ of the potential $V_1(r)$ for the intensity of laser $L_1$ being 1 W/cm$^2$.}
  \centering  
  \begin{tabular}{c c c c }
   \hline\hline
   
     $v_2$          &      $v_1$      &       $G_1$(MHz)             \\ [0.5ex]
    \hline
         
                    &      9               &     3.269                   \\
         
                    &      10              &     3.649                 \\
         
         44         &      11              &     3.176                   \\
         
                    &      15              &     2.612                    \\
         
                    &      18              &     2.374                   \\ [1ex]
      \hline
         
  \end{tabular}
  \label{table1}
\end{table}
we have calculated 71 and 48 bound states of $V_1(r)$ and $V_2(r)$ potentials, respectively.
To calculate the transition dipole elements between the continuum and the bound state or 
between the two bound states, we have used the transition dipole moment data of Ref. \cite{arpita2016}.  
 Molecular dipole transitions between two ro-vibrational states or between continuum and bound state is governed by FC principle. 
According to this principle, for excited vibrational (bound) states, bound-bound or continuum-bound transitions mainly occur near the turning points of bound states.
In general, highly excited vibrational wave functions of  diatomic molecules or molecular ions have their maximum amplitude near the outer turning points.
 Spectral intensity is proportional to the square of the FC overlap integral or FC factor.
 It implies  that PA spectral intensity will be significant when the continuum 
state has a prominent anti-node near the separation at which outer turning point of the excited  bound state lies. The value of free-bound FC factor for
 $v_2 = 44$ is found to be quite high since  the maximum of the excited bound state wave function near the outer turning point
coincides nearly with a prominent anti-node of the ground-state scattering wave function. We have also calculated the value of FC factor between the continuum of $V_1(r)$ potential
and different bound states of the $V_2(r)$ potential and it is found to be very small.

Let $\mid \phi_{v J} \rangle$ represents the ro-vibrational bound state in the potential $V_2(r)$ with 
vibrational and rotational quantum number $v$ and $J$, respectively. The energy-normalized partial-wave scattering state in the potential $V_0(r)$ is denoted by $\mid \psi_{\ell,k}(r) \rangle$ where $\ell$ is the partial wave of the relative motion between the ion and the atom; and $k$ is the wave number related to the collision energy by $E = \hbar^2 k^2 /2\mu $ with $\mu$ being the reduced mass of the ion-atom pair. The free-bound transition dipole moment element ${\mathcal D}_{v J,\ell}$ is defined by 
\bea         
{\mathcal D}_{v J,\ell} = \langle \phi_{v J}\mid {\mathbf D}(r)\cdot {\mathbf E}_{L_2}\mid \psi_{\ell,k}(r)\rangle = \mid E_{L_2}\mid ^2 \eta_{J,\ell} D_{v, \ell} \eea 
where ${\mathbf D}(r)$ is the molecular transition dipole moment that depends on the internuclear separation $r$, 
$\eta_{J,\ell} \le 1 $ is an angular factor and 
\bea 
D_{v, \ell} = \int d r \phi_{v J}(r) D(r) \psi_{\ell,k}(r)
\eea 
is the matrix element of radial part with  $D(r)$ being the transition dipole moment, $\phi_{v J}(r) = \langle r \mid \phi_{v J} \rangle$ 
and $\psi_{\ell,k}(r) = \langle r \mid \psi_{\ell,k} \rangle $ represent the bound and the scattering wave functions. The quantity $D_{v, \ell}$ is weakly dependent on $J$ and 
so we ignore the $J$-dependence of $D_{v, \ell}$. The photoassociative stimulated line width is given by $\Gamma(E) = 2 \pi \mid  \eta_{J,\ell} D_{v, \ell}\mid ^2 $. We plot the quantity  $|D_{v, \ell}|^2 $ as a function 
of energy $E$ for a number of partial waves $\ell$ up to $\ell=10$.

\begin{figure}
 \includegraphics[width = 6.0 in]{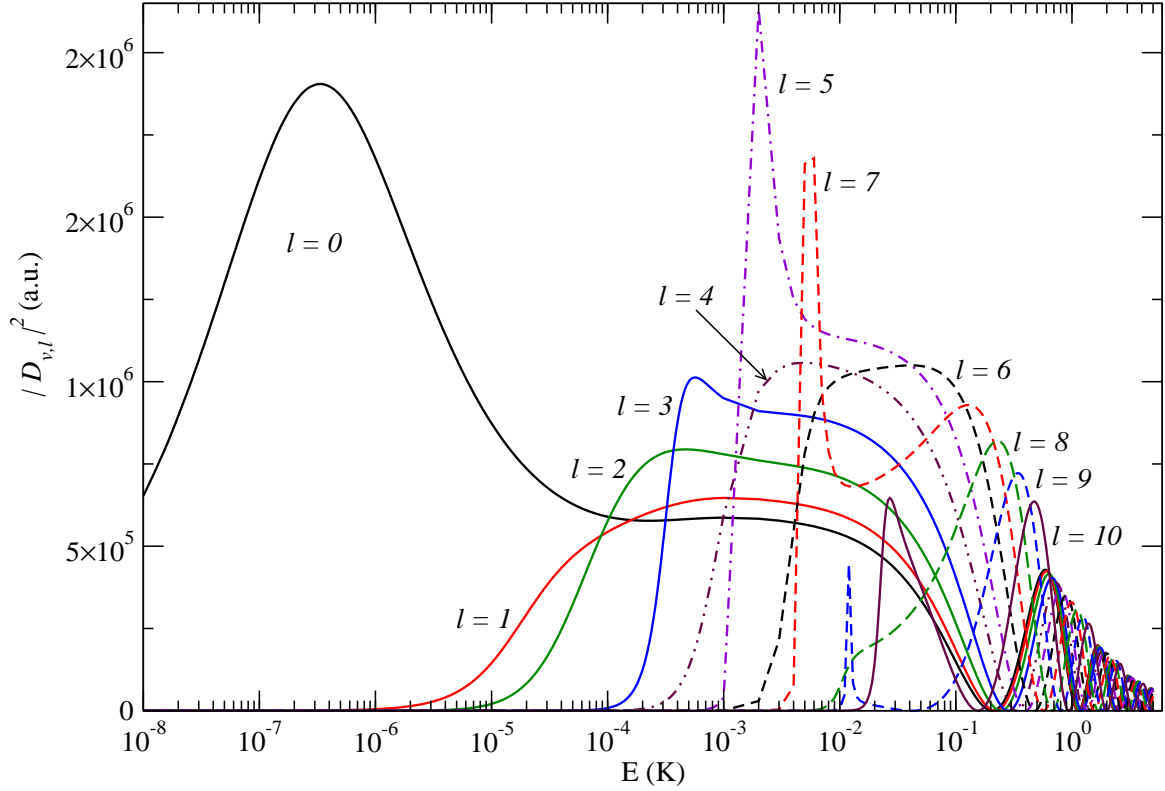}
  \caption{ The Square of free-bound radial transition dipole moment in a.u for ground continuum sates with $\ell$ ranging from 0 to 10 and 
  excited bound vibrational level with $v_2 = 44$ of $V_2(r)$ potential.}
 \label{fig3.}
\end{figure}

It is clear from FIG.\ref{fig3.} that the contributions to $|D_{v, \ell}|^2 $ of higher partial waves are larger than that of $s$-wave for the energy above 
to 0.1 mK. At very lower energy ($E  <  0.1$mK), only $s$-wave makes finite contribution to the dipole transition. The higher partial-wave 
contributions in the sub-milliKelvin or milliKelvin regime can be attributed to the relatively stronger attractive nature of the long-range part of 
ion-atom interaction which goes as $\sim - 1/r^4$ as $r \rightarrow \infty$. This is unlike that for neutral atom-atom interaction that asymptotically goes 
as $\sim - 1/r^6$. Therefore, compared to neutral atom-atom case where only a few low lying partial waves become important at such temperatures, a larger number of higher partial waves contribute 
significantly even in the sub-milliKelvin regime. Furthermore, the higher partial waves exhibit a prominent anti-node in the sub-milliKelvin or milliKelvin regime. For instance let us consider 
the contribution from $\ell = 3$. Its contribution is maximum at energy near 0.3 milliKelvin and $|D_{v, \ell=3}|^2 $ remains more or less steady for energy ranging from 0.3 milliKelvin to 10 milliKelvin, but its contribution is negligible for energy lower than 0.1 milliKelvin. 

For the sake of simplicity let us neglect the hyperfine interactions. This assumption can be well justified  for our system. As per FC principle, PA transitions predominantly occur at the separation $r$ near the outer turning point which is about 19 $a_0$ for $v_2=44$. At this separation the value of central potential $C_4/r^4$ is of the order of 1000 GHz. The hyperfine splitting in the ground-state of (LiCs$^+$) is of the order of 100 MHz. So, in comparison to the central interaction, we can safely neglect hyperfine interaction. 
Since the total molecular angular momentum is given by $\mathbf J = \mathbf S +\mathbf L +\ell$, where $\mathbf S$ and $\mathbf L$ are the total electronic spin and orbital quantum number, respectively. 
 In our model $\mathbf L = 0$ and $\mathbf S= \frac{1}{2}$ for ground-state potential $V_0(r)$.
Now, if we tune the PA laser to the near-resonance for ro-vibrational state ($v=44, J=\frac{9}{2}$) then 
only $\ell =3 $ will be coupled by the PA laser due to selection rule $\Delta J=\pm 1$. Now, if we choose $\bar{E} = 0.3 $ milliKelvin and accordingly adjust the parameters $\Delta_1$ and $\Delta_2$, then it is expected that the ion-atom pair will be 
in an approximate dark state for a range of collision energies near $\bar{E}$. For a thermal system of ion-atom mixture, we need to set the temperature $k_{b} T \simeq \bar{E}$ in order to bring 
a finite fraction of the atom-ion pairs in the approximate DS. Now, under these conditions, if we slowly vary the ratio $\beta = |\Lambda_{2 \bar{E}}|^2/\hbar G_1 $ subject to the condition of Eq.(\ref{adia}), then we can 
accomplish an almost STIRAP process for transfer of cold atom-ion pairs into cold ionic molecules. 

\begin{figure}
\includegraphics[width=6.00 in]{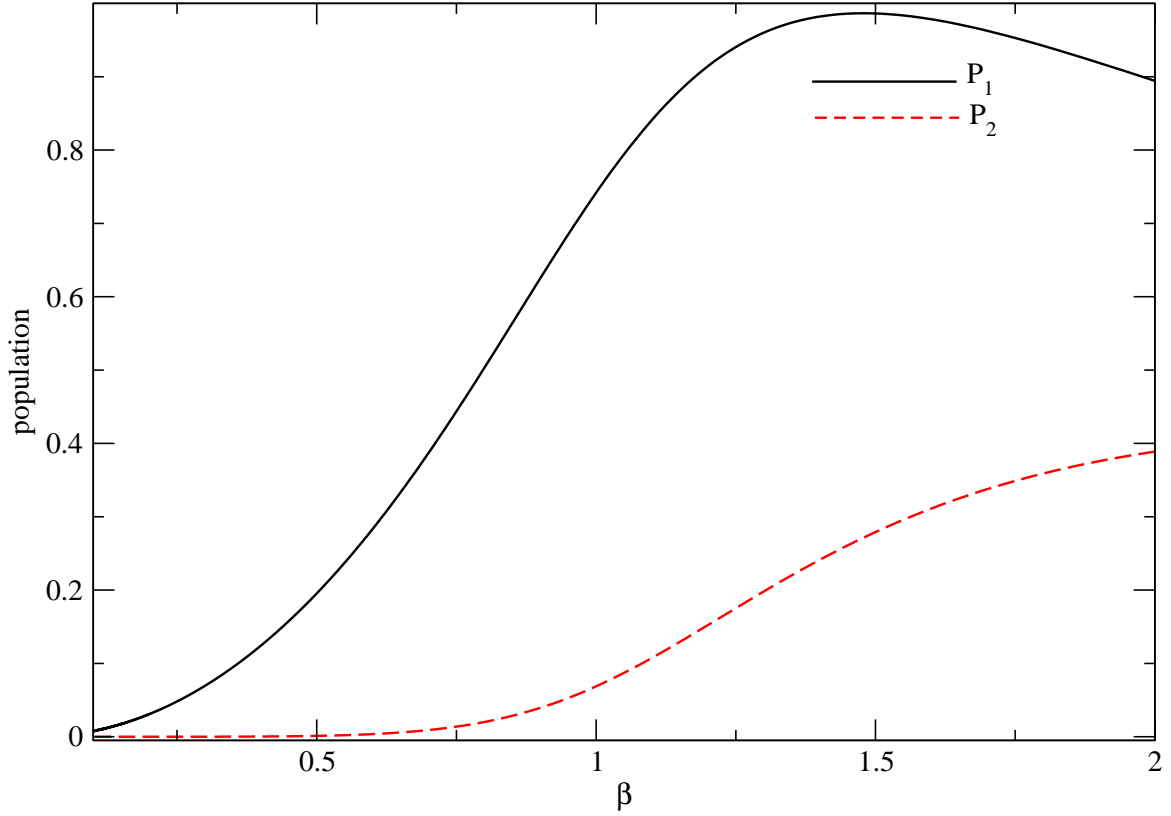}
 \caption{Solid line and dotted line  represents the population of the final and intermediate state, respectively  for (LiCs)$^+$ ionic system.}
 \label{fig4.}
\end{figure}

In FIG.\ref{fig4.}, we have shown the variation of $P_1$ and $P_2$  with $\beta$. Although, the
 final state probability approaches unity ($P_1\simeq 1  $) for all value of $\beta$, but
 there is a significant probability of the inter-mediate state ($P_2\simeq 0.4$). Hence possibility of formation of 
 dark state is very low for higher values of $\beta$. But in the lower range of $\beta$, we 
 find that the probability of intermediate state is negligibly small  
 whereas that of probability of final state rises gradually. So in this region where the value of
 $G_1$ is much more higher, the system remains almost in the dark state and population transfer efficiency is about 50 percent. Finally, 
the formation of meta-stable molecular ion takes place in $V_1(r)$ state. This meta-stable
molecular ion have longer life time. Therefore, the formation of ground-state molecular ion
becomes more likely due to favorable level structure.
To find coherent laser coupling between the two selected bound states,  we have calculated Rabi frequency $G_1$ which is expressed as 
\begin{equation}
\hbar G_1 = \Bigg(\frac{I_{L_1}}{4\pi\epsilon_0 c}\Bigg)^\frac{1}{2} \langle\phi_{v j} \mid \mathbf{D}(r)\cdot\hat{\mathbf\epsilon} \mid \phi_{v^\prime j^\prime} \rangle|
\end{equation}
where $\hat{\epsilon}$ is the unit vector of laser polarization and $\mid \phi_{v j}\rangle$ and $\mid \phi_{v^\prime j^\prime}\rangle$ are the two 
bound states. In Table.\ref{table1}, we have shown the calculated Rabi frequency between the different
bound levels of $V_1(r)$ potential and bound state ($v = 44$) of $V_2(r)$ potential for 
laser intensity $I_{L_1} = 1$ W/cm$^2$ of laser $L_1$. From the Table.\ref{table1}, the maximum Rabi frequency corresponding to the
bound-bound transition is found to be 3.649 MHz. Comparing this value with the calculated spontaneous line width  $\gamma = 26$ MHz of the excited bound state of  $V_2(r)$ potential  by using standard formula \cite{rakshit2011}, we infer that the life time of the photoassociated molecule in these 
excited molecular bound state is very small and decay spontaneously. Hence, it is difficult to form molecular ion by two-photon incoherent Raman PA.

      Now we discuss the possibility of the formation of ground-state molecular ion by three photon process.
 From the above discussion we find that the excited molecular bound states in $V_2(r)$ potential are lossy in nature. Here we  
 use a two-photon scheme by which the contribution of this lossy state is effectively minimized.
The energy dependence of $\Gamma_{2\bar E}$ arises from the FC factor. Now, taking the value of 
$\Gamma_{2\bar E} = 1.43$ MHz that corresponds to FC factor at an energy near 0.3 mK and intensity
 6 W/cm$^2$ of $L_2$. In our calculation, we vary $G_1$ from 0.01 MHz to 10 MHz which corresponds to the variation of $\beta$ from 143 to 0.143. Adiabatic condition will be maintained if the time rate of change of $\beta$ is kept much less than 6.3 per second.

\section{Conclusions}
 In conclusion, we have shown that for a continuum-bound-bound photoassociative ion-atom system resembling $\Lambda$-type configuration, creation of an approximate dark state at low energy is possible by driving stimulated Raman transitions with two lasers under certain physical conditions. By making use of this state, 
 the initial population in the continuum of ion-atom scattering states can be partially and coherently transferred to a long-lived molecular bound state via an intermediate excited bound state in a STIRAP-like fashion. We have illustrated our proposed method by numerical simulation with  a realistic model system of (LiCs)$^+$ system. We have developed our treatment based on Fano's method of diagonalisation of a continuum-bound coupled system. The approximate dark state is a dressed continuum which is an admixture of a long-lived bound state and the ground-state continuum. The dark state becomes decoupled from the lossy state only for a particular collision energy. Therefore, complete population transfer is, in general, not possible in a continuum-bound system in thermal equilibrium. However, by a proper choice of temperature, laser detunings and rotational levels  of the lossy state, one can optimize the population transfer. In this paper, we have considered a bare continuum. As a further study, one can 
consider two-photon Raman process involving a magnetic Feshbach-resonance induced structured continuum \cite{deb_pra_2014}. In such physical situations, it is possible to create a bound state in continuum that can be effectively decoupled the continuum \cite{deb_pra_2012}. This opens up the possibility of creating an effective $\Lambda$ system of three bound states involving an underlying continuum. Whether  complete or more efficient coherent population transfer in such an effective $\Lambda$ system  is possible or not would be an interesting problem to pursue in future. 
 \ack
D. Sardar is grateful to CSIR, Government of India, for a support. 
This work is also jointly supported by Department of Science and Technology (DST), Ministry of Science and Technology, Govt. of India and Ministry of Higher Education and Scientific Research (MHESR), Govt. of Tunisia, under an India-Tunisia Project for Bilateral Scientific Cooperation.
 
 \appendix{}
\section{Dressed continuum }

We diagonalise the hamiltonian $H$ given in Eq.(\ref{h0}) and Eq.(\ref{hi}) in two steps using Fano's method. First, we diagonalise the sub-system comprising the bare continuum of states $\mid E\rangle$ and the bound state $\mid 2\rangle$ which are coupled by laser pulse $L_2$. At this stage, we define an intermediate dressed state $\mid E\rangle_{idr}$  as
 \begin{eqnarray}
\mid E\rangle_{idr} &=& A_{2E}\mid2\rangle + \int C_E(E') \mid E'\rangle dE'  
\label{idrs_state}                                
\end{eqnarray}
where $A_{2E} = \frac{\Lambda_{2E}}{E+\hbar\Delta_2-E_1^{sh}-i\hbar\Gamma_{2E}/2}$, $C_E(E') = \frac{A_{2E} \Lambda_{E'2}}{E-E'} + \delta(E-E')$ and $\Gamma_{2E} = \frac{2\pi |\Lambda_{2E}|^2}{\hbar}$ is the energy-dependent stimulated line width for the free-bound PA transition. $E^{sh}_{1} = {\cal P}\int \frac{| \Lambda_{2E'}|^2 dE'}{E - E'}$ is the light shift of the bound state, where ${\cal P}$ represents the Cauchy Principle value integral. In terms of $\mid E\rangle_{idr}$, the hamiltonian can be written as 
\begin{eqnarray}
\label{hdr1}
\hspace{-1cm} H = \hbar\left(\Delta_1 - \Delta_2\right)\mid 1\rangle\langle 1\mid + \int\mid E\rangle_{idr \hspace{1mm} idr}\langle E\mid dE +      
\Bigg(\hbar G_1 |1\rangle\langle2| + {\rm H.c.}\Bigg)
\end{eqnarray}
Using Eq.(\ref{idrs_state}) and the completeness relation
\begin{eqnarray}
\mid 1\rangle\langle1\mid+\int\mid E\rangle_{idr \hspace{1mm} idr}\langle E\mid dE=1
\end{eqnarray}
we can write 
\begin{eqnarray}
\label{2dr1}
\mid 2\rangle=\left(\mid 1\rangle\langle1\mid+\int\mid E\rangle_{idr \hspace{1mm} idr}\langle E\mid dE\right)\mid 2\rangle=\int A_{2E}^*|E\rangle_{idr} dE
\end{eqnarray}
Substituting Eq.(\ref{2dr1}) into Eq.(\ref{hdr1}) we get 
\begin{eqnarray}
\label{hdr1f}
H &=& \hbar\left(\Delta_1 - \Delta_2\right)\mid 1\rangle\langle 1\mid + \int\mid E\rangle_{idr \hspace{1mm} idr}\langle E\mid dE\nonumber\\ &+&      
        \left(\int\Lambda_{E1} \mid E\rangle_{idr}\langle1 \mid dE + {\rm H.c.}\right) 
\end{eqnarray}
The above Hamiltonian describes the one bound state $\mid 1\rangle$  coupled to $\mid E\rangle_{idr}$ by an effective coupling parameter $\Lambda_{1E}=\hbar G_1 A_{2E}$. In the second and final step we diagonalise \ref{hdr1f}. We define the final dressed state as
 \begin{eqnarray}
\hspace{-1cm}\mid E\rangle_{dr} &=& B_{1E}\mid1\rangle + \int D_{E}(E'') \mid E''\rangle_{idr} dE''\nonumber\\  &=& B_{1E}\mid1\rangle + \int D_{E}(E'') \left(A_{2E''}\mid2\rangle + \int C_{E''}(E') \mid E'\rangle dE'\right) dE''\nonumber\\&=&B_{1E} \mid 1\rangle + B_{2E} \mid 2\rangle + \int F_E(E') \mid E'\rangle
\label{drsf_state}                                
\end{eqnarray}
The expression for $B_{1E}$ is
 \begin{eqnarray}
B_{1E} = \frac{\Lambda_{1E}}{E+\hbar(\Delta_2-\Delta_1)- E_2^{sh}-i\hbar\Gamma_{1E}/2}  
 \end{eqnarray}
where $\Gamma_{1E}=2\pi\hbar|G_1|^2|A_{2E}|^2$ is the effective stimulated linewidth of $\mid1\rangle $ and $E^{sh}_{2}= {\cal P} \int \frac{| \Lambda_{2E}|^2 dE}{\left[(E+\hbar\Delta_2 - E^{sh}_1)^2 + \hbar^2 \Gamma^{2}_{2E}/4\right](E-E')}$ is the laser induced shift of the same. Using the expression of $A_{2E}$ as given above, the explicit form of $B_{1E}$ is
 \begin{eqnarray}
 \label{b1e}
\hspace{-1.5cm}B_{1E} = \frac{\hbar G_1\Lambda_{2E}\left(E+\hbar\Delta_2-E_1^{sh}+i\hbar\Gamma_{2E}/2\right)}{\left[E+\hbar(\Delta_2-\Delta_1)- E_2^{sh}\right]\left[\left(E+\hbar\Delta_2-E_1^{sh}\right)^2+\frac{\hbar^2\Gamma_{2E}^2}{4}\right]-\frac{i}{2}\hbar^3G_1^2\Gamma_{2E}} 
 \end{eqnarray}
The other terms are $D_{E}(E'')= \frac{B_{1E} \Lambda_{E''1}}{E-E''} + \delta(E-E'')$ and 
 $B_{2E} = \int D_{E}(E'') A_{2E''} dE''$. The explicit form of $B_{2E}$ can be written as 
\begin{eqnarray}
 \label{b2e}
\hspace{-2.5cm} B_{2E} =\frac{\left[E+\hbar(\Delta_2-\Delta_1)\right]\left[E+\hbar \Delta_2 -E^{sh}_1 -i\hbar\Gamma_{2E}/2\right] \Lambda_{2E}}{\left[E+\hbar(\Delta_2-\Delta_1)-E^{sh}_2\right]\left[(E+\hbar\Delta_2-E^{sh}_1)^2+(\hbar\Gamma_{2E}/2)^2\right] +i\hbar^3\Gamma_{2E} G^2_1/2 }.                 
\end{eqnarray}
Here $F_E(E')=\int D_{E}(E'') C_{E''}(E')dE''$.
Using Eq.(\ref{b1e}) and Eq.(\ref{b2e}) we finally obtain the expressions for $P_{1}$ and $P_{2}$ as
\begin{eqnarray}                                      
 \hspace{-1.5cm} P_1=\int | B_{1E}|^2 dE = \frac{\hbar^3}{2} |G_1|^2\int Z^{-1}\Gamma_{2E}(E+\hbar \Delta_2-E^{sh}_1)^2 dE \nonumber \\
  +\frac{\hbar^5}{8} |G_1|^2\int Z^{-1} \Gamma_{2E}^3 \hspace{0.1 cm}  dE 
\end{eqnarray}
and
\begin{eqnarray}
 \hspace{-1.5cm}P_2=\int | B_{2E}|^2 dE =\frac{\hbar}{2}\int  Z^{-1}  [E+\hbar(\Delta_2-\Delta_1)]^2 (E+\hbar \Delta_2 -E^{sh}_1)^2\Gamma_{2E}dE \nonumber \\ 
                                       + \frac{\hbar^3}{8} \int Z^{-1}[E+\hbar(\Delta_2-\Delta_1)]^2 \Gamma_{2E}^3 dE  
\end{eqnarray}
where 
\begin{eqnarray}
\hspace{-1.5cm}Z=[E+\hbar(\Delta_2-\Delta_1)-E^{sh}_2]^2[(E+\hbar\Delta_2-E^{sh}_1)^2+(\hbar\Gamma_{2E}/2)^2]^2 \nonumber \\ 
         \hbar^6\Gamma^2_{2E} G^4_1/4] 
\end{eqnarray}

\end{document}